\documentclass[twocolumn,showpacs,preprintnumbers,amsmath,amssymb]{revtex4}

\usepackage{graphicx}
\usepackage{dcolumn}
\usepackage{bm}
\usepackage{epsfig}
\usepackage{color}

\begin{document}


\title{Modeling the nonlinear dielectric response of glass formers}

\author{U. Buchenau}
 \email{buchenau-juelich@t-online.de}
\affiliation{%
J\"ulich Center for Neutron Science, Forschungszentrum J\"ulich\\
Postfach 1913, D--52425 J\"ulich, Federal Republic of Germany
}%

\date{May 18, 2017}

\begin{abstract}
The recently developed pragmatical model of asymmetric double-well potentials with a finite lifetime is applied to nonlinear dielectric data in polar undercooled liquids. The viscous effects from the finite lifetime provide a crossover from the cooperative jumps of many molecules at short times to the motion of statistically independent molecules at long times. The model allows to determine the size of cooperatively rearranging regions from nonlinear $\omega$-data and throws a new light on a known inconsistency between nonlinear $\omega$ and $3\omega$-signals for glycerol and propylene carbonate.
 \end{abstract}

\maketitle

\section{Introduction}

Most of our knowledge about the cooperativity of the flow process in undercooled liquids stems from numerical simulations \cite{cavagna,bb}. But numerical simulations have the weakness that they study the flow process close to the vibrational frequency regime. To study the cooperativity over a large dynamical range, one needs real experiments. The most promising of these are measurements of the nonlinear dielectric effect \cite{weinstein,weinstein2,wang,huang,tca,brun,brun2,brun3,bauer3,bauer1,view,samanta}.

The importance of the nonlinear dielectric effect for the understanding of the flow process in undercooled liquids has been gradually established within the last decade by the pioneering work of the Richert group \cite{weinstein,weinstein2,wang,huang,tca,samanta,r2016,chi5}. In a strong alternating electric field, one finds a quadratic increase of the imaginary part $\epsilon''(\omega)$ of the complex dielectric susceptibility $\epsilon(\omega)$ with increasing strength of the electric field on the high-frequency side of the $\alpha$-peak, while the low-frequency side remains essentially unchanged. Thus there is a net positive effect in the imaginary part. On the other hand, the zero frequency dielectric constant $\epsilon_s$ decreases slightly \cite{weinstein2}, consistent with the Langevin saturation for the molecular dipole moment. 

A further important experimental information was obtained from the study of the transient behavior of the nonlinear dielectric response after switching on or switching off a strong alternating electric field \cite{huang}. It was found that the effect developed its full strength within a relaxation time which was considerably shorter than the terminal relaxation time at the given temperature, consistent with a whole heterogeneous ensemble of relaxational modes of which each reacts with its own relaxation time.

In the first systematic study of the effect in glycerol \cite{weinstein}, the nonlinearity was attributed to a storage of energy in the slow relaxational modes of the flow process. In this explanation, the energy input of the oscillating electric field does not go into the phonon bath, but is stored into the slow relaxational degrees of freedom of the $\alpha$-process according to their contribution to the dynamic heat capacity. This heat input shifts the slow degree of freedom to a higher effective temperature, where it is faster and stronger. The mechanism is supposed to be the same as in hole burning \cite{schiener}.

The explanation, often denoted as "box model", has two advantages: (i) it explains why each mode reacts with its own relaxation time (ii) it provides a quantitative explanation without free parameters (one only adapts the configurational fraction of the excess entropy to a reasonable value). But it has the disadvantage that one has to postulate an energy storage mechanism for a local structural rearrangement, a hard task for the physical intuition.

The first systematic measurement of the $3\omega$ signal in glycerol \cite{brun,brun2,brun3}, recently corroborated by a more extensive study in several glass formers \cite{bauer3,view} proposed an alternative explanation in terms of extended cooperative rearrangements. The rearrangements were taken to be thermally activated jumps in asymmetric double-well potentials \cite{brun3,gregor}. But this explanation has not yet grown into a full quantitative description, neither the one of the effect itself nor the one of the transient behavior.

The present paper intends to do a first step to fill this gap. In order to do this properly, it turns out to be necessary to take the finite lifetime of the the asymmetric double-well potentials into account, using a recent pragmatical model \cite{visc,visco} for the crossover from the back-and-forth-jumps to the no-return jumps of the viscous flow. The viscous effects provide the crossover from cooperative rearrangements at short times to the motion of independent molecules at long times.

The paper is organized as follows: After this introduction, Section II presents the theoretical description. Section III contains the comparison to experimental data. Section IV discusses the results and concludes the paper.

\section{Theoretical description}

\subsection{Low frequency limit: the dipole gas}

In an undercooled liquid consisting of polar molecules, the equilibrium polarization in a strong electric field $E$ is known both from theory \cite{vanvleck} and from experiment \cite{weinstein2} to follow the Langevin saturation function
\begin{equation}\label{langevin}
  \frac{P(a_1)}{N\mu}=\frac{\coth{a_1}}{a_1}=\frac{a_1}{3}-\frac{a_1^3}{45}+\frac{2a_1^5}{945}-..,
\end{equation}
where $a_1=E\mu/kT$, $\mu$ is the dipole moment of a single molecule, and $N$ is the number density of the molecules.

The Langevin function is derived from the theoretical picture of a noninteracting gas of dipoles, each of which samples all possible orientations without caring about the orientation of the others.

This idealization is certainly not true in real polar liquids. Nevertheless, the Langevin function remains a good description. In fact, it contains a free parameter which allows to describe the effects of the interaction to first order: the effective dipole moment of a single molecule (sometimes expressed in terms of the Kirkwood correlation factor \cite{kirkwood,cole}), which is temperature-dependent in the presence of an interaction.

There is an additional factor to be taken into account, because the field $E$ seen by the molecule is not the externally applied field $E_e$. The field $E$ within the sample is larger than the external field $E_e$ by the Onsager factor \cite{weinstein2,vanvleck,onsager,kirkwood,cole}
\begin{equation}\label{On}
	f_{On}=\frac{\epsilon_s(n^2+2)^2}{3(2\epsilon_s+n^2)},
\end{equation}
where $n$ is the refractive index describing the electronic polarization of the sample and $\epsilon_s$ is the low frequency limit of the dielectric constant.

Linear response data do only allow to determine the product of the square of the effective dipole moment $\mu$ and the Onsager factor. In principle, the combination of linear and nonlinear data enable a separate determination. In fact, the validity of the Onsager relation, eq. (\ref{On}), has already been demonstrated along this line of thought from nonlinear dielectric data in propylene glycol \cite{weinstein2}.

As pointed out by one of the Referees of this paper, the Kirkwood correlation factor, which depends on the interaction of a molecule with its neighbors \cite{kirkwood,cole,booth,fulton}, can be quite different in linear and nonlinear effects. 

Let $R_s$ be the linear Kirkwood correlation factor, $R_p$ the one for the third order effect \cite{parry,piekara} and $R_5$ the one for the fifth order. Then the Langevin equation changes to
\begin{equation}\label{langevin1}
  \frac{P(a_1)}{NR_s\mu}=\frac{a_1}{3}-f_3\frac{a_1^3}{45}+f_5\frac{2a_1^5}{945}-...,
\end{equation}
with $f_3=R_p/R_s^3$, $f_5=R_5/R_s^5$, and
\begin{equation}
	a_1=\frac{R_sf_{On}E_e\mu}{kT}.
\end{equation}
$R_s\mu$ is the temperature-dependent effective molecular dipole moment, in the following again simply denoted by $\mu$.

In propylene glycol \cite{weinstein2}, $f_3$ is close to 1. In water, methanol, ethanol, propanol and butanol \cite{parry}, $f_3$ is 1.24, 1.9, 1.4, 1.5 and 1.1, respectively.

These values were determined from the Piekara factor $A=\Delta\epsilon_s/E_e^2$, where $\Delta\epsilon_s$ is the nonlinear change of the static susceptibility $\epsilon_s$, assuming the validity of eq. (\ref{On}) for the Onsager factor.  

At low enough frequency, the polarization in an alternating field $E\cos{\omega t}$ follows the field instantaneously. Thus one has up to third order
\begin{align}
	\nonumber \frac{P(t)}{N\mu}=\frac{a_1\cos{\omega t}}{3}-f_3\frac{a_1^3(\cos{\omega t})^3}{45}\\ 
	=\left(\frac{a_1}{3}-f_3\frac{a_1^3}{60}\right)\cos{\omega t}-f_3\frac{a_1^3}{180}\cos{3\omega t}.
\end{align}
One sees that there are two third-order effects: the response at the frequency $\omega$ is reduced by $f_3a_1^3/60$ and there is a signal at $3\omega$.

The complex susceptibility $\chi$ is defined by $P/\epsilon_0E_e$ and is given up to third order by
\begin{equation}
\chi(\omega)=\frac{P}{\epsilon_0E_e}=\chi_1(\omega)+\frac{3\chi_3^{(1)}(\omega)}{4}E_e^2+\frac{\chi_3^{(3)}(\omega)}{4}E_e^2,
\end{equation}
where $\chi_1(\omega)$ is the linear response susceptibility. $\chi_3^{(1)}(\omega)$ and $\chi_3^{(3)}$ describe the third order response at $\omega$ and $3\omega$, respectively. 

Our consideration shows that the low frequency limit is described up to third order in the external field $E_e$ by three real dimensionless numbers $\Delta\chi$, $\chi_3^{(1)}(0)E_e^2/\Delta\chi$ and $\chi_3^{[3)}(0)E_e^2/\Delta\chi$ with

\begin{equation}\label{deltachi1}
	\Delta\chi=\frac{N\mu^2f_{On}}{3\epsilon_0kT}
\end{equation}
\begin{equation}\label{chi31}
	\frac{\chi_3^{(1)}(0)E_e^2}{\Delta\chi}=-f_3\frac{a_1^2}{15}
\end{equation}
\begin{equation}\label{chi33}
	\frac{\chi_3^{(3)}(0)E_e^2}{\Delta\chi}=-f_3\frac{a_1^2}{15}.
\end{equation}

$\Delta\chi$ is the difference between the dielectric constants $\epsilon_s$ at very low and $\epsilon_\infty=n^2$ at very high frequency.

The Piekara factor
\begin{equation}
	\frac{\Delta\epsilon_s}{E_e^2}=-\frac{\Delta\chi}{E_e^2}f_3\frac{a_1^2}{15},
\end{equation}
which shows that the static nonlinear effect fixes the low frequency limits for both $\chi_3^{(1)}(\omega)$ and $\chi_3^{(1)}(\omega)$.

For the $3\omega$-signal, there is a second useful definition of a dimensionless susceptibility \cite{brun,brun2}
\begin{equation}\label{x3}
	X_3(\omega)=\frac{NkT}{\epsilon_0\Delta\chi^2}\left|\chi_3^{(3)}(\omega)\right|.
\end{equation}
Inserting eq. (\ref{deltachi1}) and eq. (\ref{chi33}) into this definition, one finds the low frequency limit of $X_3(\omega)$
\begin{equation}\label{x30}
	X_3(0)=\frac{f_3f_{On}}{5}.
\end{equation}

In the same way, one can define a dimensionless measure of the $5\omega$-signal
\begin{equation}
	X_5(\omega)=\frac{N^2k^2T^2}{\epsilon_0^2
	\Delta\chi^3}\left|\chi_5^{(5)}(\omega)\right|
\end{equation}
and gets
\begin{equation}\label{x50}
	X_5(0)=\frac{2f_5f_{On}^2}{7}.
\end{equation}

Equs. (\ref{x30}) and (\ref{x50}) are in direct contradiction to data \cite{brun2,bauer3,r2016} at $3\omega$ and to new $5\omega$-data \cite{chi5} in glycerol and propylene carbonate, because there the Onsager factors are missing. We will come back to this question in the comparison to experiment in Section III and in the discussion in Section IV.  

\subsection{Cooperatively rearranging regions}

While the low frequency behavior of an undercooled liquid of polar molecules is understandable in terms of independent molecular orientations, this is no longer true at higher frequencies. At short times the undercooled liquid is like a glass, with thermally activated structural rearrangements from one minimum of the energy landscape to another. In such a rearrangement, a large number of molecules changes its orientation. The change of orientation is given by the requirement that the new structure should be again a stable one. 

Let us consider a region of $N_{corr}$ molecules which undergoes a thermally activated jump into another stable structure. As recently pointed out \cite{brun3}, the mean square of the dipole moment change $\Delta\mu$ of the whole region must be expected to be a factor $N_{corr}$ larger than the average mean square dipole moment change of a single molecule, increasing the corresponding nonlinear dielectric effect.

A thermally activated jump rate is proportional to $\exp(-V/kT)$, where $V$ is the energy difference between the structural energy minimum and the saddle point on the way to the next structural energy minimum. An electric field $E$ changes this rate by the factor $\exp(E\Delta\mu/2kT)$. Here $\Delta\mu$ is the dipole moment difference of the two energy minima of the region in field direction.
\begin{equation}\label{a}
	a=\frac{E\Delta\mu}{kT}
\end{equation}
If one reaches one half of the dipole momentum change of the whole jump at the saddle point, the rate changes up to third order by the factor \cite{brun3}
\begin{equation}\label{rate}
	1+\frac{a}{2}+\frac{a^2}{8}+\frac{a^3}{48}
\end{equation}
where the linear term gives rise to the linear dielectric response and the two higher powers are responsible for the nonlinear dielectric effect.

But it is not only the rate change which gives rise to nonlinearity; it is also the population change of the two wells in the strong alternating field. As shown independently by Diezemann \cite{gregor} and the french group \cite{brun3}, the asymmetry $\Delta$ (the energy difference of the two structural energy minima separated by the barrier $V$) is crucial for the nonlinear effects. The asymmetry $\Delta=1.31695 kT$ (corresponding to $(\tanh(\Delta/2kT))^2=1/3$) divides two regions with different behavior. For smaller asymmetries, the low frequency real part of the nonlinear effects is negative, for higher asymmetries positive.

According to Diezemann \cite{gregor}, the complex amplitude $p$ of the dipole moment of the asymmetric double-well potential in the strong field $E\cos{\omega t}$ is given by
\begin{equation}\label{greg}
	p=\frac{2a\Delta\mu}{3(\cosh{\delta})^2}\left(\frac{1}{1+i\omega\tau}+\frac{3a^2}{20}(S_3^{(1)}+S_3^{(3)})\right),
\end{equation}
where $\delta=\Delta/2kT$ and the two nonlinear terms, $S_3^{(1)}$ and $S_3^{(3)}$, are given by
\begin{align} \label{s31}
\nonumber S_3^{(1)}=(\tanh{\delta})^2\frac{3(1-2ix)}{(1+x^2)(1+4x^2)}\\
+\frac{2(x^2-1)-ix(x^2-3)}{2(1+x^2)^2}	
\end{align} 
for the term on the frequency $\omega$ and
\begin{align} \label{s33}
\nonumber S_3^{(3)}=(\tanh{\delta})^2\frac{(1-11x^2)-6ix(1-x^2)}{(1+x^2)(1+4x^2)(1+9x^2)}\\
+\frac{2(5x^2-1)-3ix(x^2-3)}{6(1+x^2)(1+9x^2)}	
\end{align}
for the term on the frequency $3\omega$, with $x=\omega\tau$. The relaxation time $\tau_r$ is given by
\begin{equation}
	\tau_r=\frac{\tau_0\exp(V/kT)}{\cosh{\delta}},
\end{equation}
with $\tau_0\approx 10^{-13}$ s.

In a real sample, one has to reckon with a whole distribution of different asymmetries. Summing up all their contributions, one obtains again equs. (\ref{s31}) and (\ref{s33}) with an average $d_2=\overline{\tanh{\delta}^2}$, supposed to be the same for all retardation processes.

Note that $d_2$ corresponds to the parameter $\delta$ in reference \cite{brun3}, which appeared simultaneously with reference \cite{gregor}.

For a constant asymmetry density, and weighting the asymmetries with the prefactor $1/\cosh{\delta}^2$ (the weakening factor for the relaxation strength of an asymmetric double well potential), one obtains the value $d_2=1/3$. 

\subsection{Nonlinear response in the pragmatical model}

The nonlinear dielectric effects are very sensitive to cooperativity, because their strength depends quadratically on the local dipole moment change of a given relaxation process. If a whole region of hundred molecules changes its structure, the dipole moment change will be larger than the one of a rotational jump of a single molecule.

In an undercooled liquid, one has both situations: At short times, the decay of the rigidity begins with local structural changes within a rigid solid, Eshelby transformations \cite{eshelby} of ten to hundred molecules. Later, the combined effect of these local changes leads to viscous flow, with an independent small-angle diffuse motion of each molecule.

To calculate the nonlinear dielectric effects, one needs a quantitative description for this crossover. Such a quantitative description is provided by the pragmatical model \cite{visc,visco}.

In its present version \cite{visco}, the pragmatical model describes the normalized dielectric relaxation function
\begin{equation}
	\Phi(\omega)=\frac{\epsilon(\omega)-\epsilon_\infty}{\epsilon_s-\epsilon_\infty}
\end{equation}
($\epsilon_s$ dielectric susceptibility at frequency zero, $\epsilon_\infty$ high frequency limit) in terms of three components.

The earliest fraction $f_r(1-f_c)$ of the response is the retardational one (back-and-forth jumps), responsible for the Kohlrausch $t^\beta$ rise at short times, with a Kohlrausch $\beta$ of about 1/2. In the simplest case
\begin{equation}\label{simple}
	l(\tau)=l_0\frac{\tau_c-\tau}{\tau_c}\left(\frac{\tau}{\tau_c-\tau}\right)^\beta.
\end{equation}
Here $l(\tau)$ is the density of Debye relaxation processes in $\ln\tau$, the logarithm of the relaxation time, the prefactor $l_0$ is determined by $f_r$ and $\tau_c$ is the structural lifetime.

The rest of the response is viscous response at the structural lifetime $\tau_c$. It consists of two components, a larger one and a smaller one. The smaller one is directly due to the same structural rearrangements, namely to those which make just a single jump within the structural lifetime, never jumping back.

To this one has to add a second and larger viscous response, due to the collective effect of all local structural rearrangements which cause the finite structural lifetime. The existence of this term has been inferred from considerations on the viscosity \cite{visco}. It is responsible for a fraction $f_c$ of the total response and is a collective small-angle motion, in which each molecule moves on its own.

In the simplest case of eq. (\ref{simple}), one has the relation
\begin{equation}\label{fr}
	f_r=1-\beta.
\end{equation}
For the derivation and a more detailed explanation, the reader is referred to the original paper \cite{visco}.

The substances of the present paper are not the simplest case. Glycerol and propylene carbonate have an excess wing at high frequencies, which can be described by the replacement 
\begin{equation}\label{excess}
	\left(\frac{\tau}{\tau_c-\tau}\right)^\beta\ \rightarrow \left(\frac{\tau}{\tau_c-\tau}\right)^\beta+f_{1/6}\left(\frac{\tau}{\tau_c-\tau}\right)^{1/6}
\end{equation}
with a small coefficient of $f_{1/6}$ of the order of 0.05. These are two contributions with different $\beta$, so $f_r$ is the sum of two contributions calculated with eq. (\ref{fr}).

The terminal relaxation time is not sharp, but varies from region to region. One describes this by a gaussian distribution of terminal relaxation times in $\ln\tau$, with a full width $W_r$ at half maximum of about 2 (a bit less than a decade). In fact, otherwise the peaks in the imaginary part of $\epsilon(\omega)$ get too sharp.

The linear response of glycerol and propylene carbonate is described by the six parameters $f_c$, $\beta$, $\tau_c$, $W_r$, $f_{1/6}$ and $\Delta\chi=\chi_s-\chi_\infty$, which are tabulated in Table I.

In the mono-alcohols \cite{bo}, the main dielectric polarization decay occurs in a Debye contribution at $\tau_{D}$, about two decades later than the structural lifetime $\tau_c$, because the molecules form long-lived chains or rings \cite{mono}. In this case, the retardation fraction $f_r(1-f_c)$ gets smaller by a factor of the order $\tau_c/\tau_D$, more precisely by the factor $f_D\tau_c/\tau_D$.

Another addition to the retardational relaxation density in the two mono-alcohol examples is the secondary relaxation peak, describable in terms of a Cole-Cole contribution
\begin{equation}
	\frac{a_{CC}}{1+(i\omega\tau_{CC})^\delta}
\end{equation}
with a small amplitude $a_{CC}$, a thermally activated $\tau_{CC}$ and a Cole-Cole exponent $\delta$ of about 0.25. Again, the concomitant change of $f_r$ is easy to calculate.

Proceeding to the nonlinear response in the pragmatical model, the retardation processes are assumed here to occur in asymmetric double-well potentials. The nonlinear response of an asymmetric double-well potential is given by Diezemanns equations (\ref{greg}), (\ref{s31}) and (\ref{s33}) in Section II. B.

In Table II, the strength of the nonlinear response of a structural rearrangement in the neighborhood of the terminal relaxation time $\tau_c$ is characterized by the ratio $\Delta\mu/\mu$ (the dipole moment change of the rearrangement in units of the dipole moment of a single molecule). 

At higher frequencies, it turns out to be necessary to weaken the nonlinear response of the structural rearrangements by the factor $1/(\omega\tau_c)^\gamma$, with a $\gamma$ of about 0.2. 

As argued in the preceding subsection, the asymmetry parameter $d_2$ is of the order of one third, where the zero frequency contribution of Diezemanns equations begins to get positive.

But at the frequency zero, one should no longer see the effect of these structural rearrangements; one should see only the single molecule effects. In order to fulfill this condition, the viscous contribution from the structural rearrangements (the one with the fraction $(1-f_c)(1-f_r)$) should neutralize the retardation contribution.

This can be done (again in a pragmatical way) by describing the viscous structural rearrangement component $(1-f_c)(1-f_r)$ in terms of Diezemanns equations for an asymmetric double-well (with the same strength $a$ as the retardation ones) at the relaxation time $\tau_c$, choosing its asymmetry parameter $d_\eta$ in such a way that one gets the desired result
\begin{equation}\label{deta} 
d_\eta=\frac{1}{3}-F_\gamma\frac{f_r}{1-f_c-f_r}\left(\frac{1}{3}-d_2\right)-\frac{f_3a_1^2}{3a^2},
\end{equation}
where the factor $F_\gamma$ of about 0.8 comes from the weakening $1/(\omega\tau_c)^\gamma$ at higher frequencies.

With this recipe, the whole ensemble of structural rearrangements comes down to the nonlinear single molecule result at zero frequency.

The collective response fraction $f_c$ is described in terms of a Debye relaxation at $\tau_c$ in a symmetric double-well potential with the dipole moment change $\Delta\mu=\mu\sqrt{f_3/3}$, which again supplies the correct zero-frequency behavior.

Finally, the dominating Debye component in the mono-alcohols is described in terms of an asymmetric double-well potential at $\tau_D$ and the dipole moment change $\Delta\mu=\mu\sqrt{F_Df_3/3}$. The asymmetry $d_{2D}$ is chosen according to
\begin{equation}
	d_{2D}=\frac{1}{3}-\frac{1}{3F_D}
\end{equation}
to insure the correct zero-frequency limit.

With these definitions, the nonlinear dielectric response of glycerol and propylene carbonate is characterized by the parameters $\Delta\mu/\mu$, the average asymmetry $d_2$, the decay parameter $\gamma$ to higher frequencies and the deviation $f_3$ from the perfect Langevin behavior.

In the mono-alcohols, one has to add the nonlinearity strength $F_D$ of the Debye line. 

\begin{figure}[b]
\hspace{-0cm} \vspace{0cm} \epsfig{file=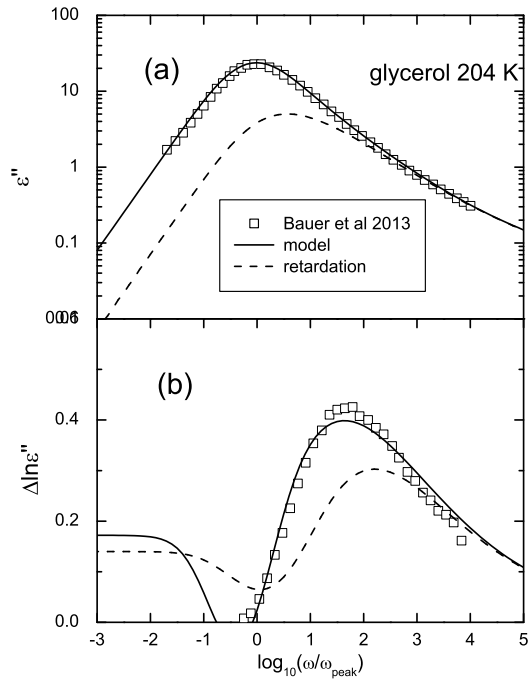,width=7 cm,angle=0} \vspace{0cm} \caption{Comparison of the model to glycerol data at 204 K (a) linear dielectric response \cite{bauer1} (b) nonlinear dielectric data at the frequency $\omega$ \cite{bauer1}.}
\end{figure}

\section{Comparison to experiment}

\begin{table}[htbp]
	\centering
		\begin{tabular}{|c|c|c|c|c|c|c|c|c|}
\hline
substance                  & $T$  & $\tau_D/\tau_c$  & $\Delta\chi$    & $\tau_c$     & $\beta$ &$f_{1/6}$ & $W_r$ &  $f_c$\\
\hline   
                           & $K$  &        &                 & $s$          &         &          &       &       \\
\hline                                                   
glyc\cite{bauer1}      & 204  &  1     & 62.7                & 0.020        &  0.55   & 0.035    &  1.6  & 0.44  \\
PC\cite{tca}           & 166  &  1     & 128.1               & 0.0056       &  0.5    & 0.041    &  1.6  & 0.65  \\
\hline
                       &      &        &                     &              &         & $f_D$    &       &       \\
                       \hline
2-EH \cite{bauer3}     & 173  &  448   &  27.7               & 0.002        &  0.42   & 8.6      & 3.9   & 0.0   \\
prop\cite{bauer2015}   & 114  & 13.2   &  73.6               & 0.013        &  0.38   & 0.66     & 1.5   & 0.0   \\
\hline		
		\end{tabular}
	\caption{Fit parameters for the linear dielectric response. Glyc is glycerol, PC is propylene carbonate, 2-EH is 2-ethyl-1-hexanol, prop is 1-propanol. The mono-alcohols 2-EH and prop have an additional secondary relaxation peak (parameters given in the text).}
	\label{tab1:lin}
\end{table}

\begin{table}[htbp]
	\centering
		\begin{tabular}{|c|c|c|c|c|c|c|c|c|}
\hline
substance                  & $T$  & $E_e$ &  $\mu$ &$\Delta\mu/\mu$& $N$           &  $f_3$  & $d_2$ & $F_D$    \\
\hline   
                           & $K$  &$kV/cm$& $D$    &               & $10^{28}/m^3$ &         &       &          \\
\hline                                                                                          
glyc \cite{bauer1}         & 204  & 671   & 4.26   &  5.1          & 0.845         &   0.72  &  .42  &          \\
PC \cite{tca}              & 166  & 177   & 5.67   &  8.1          & 0.796         &    1*   &  .35  &          \\
\hline
prop \cite{bauer2015}      & 114  & 468   & 2.47   &  0.66         & 1.020         &    1.81 &  .36* & 1.1      \\
2-EH \cite{bauer3}         & 173  & 460   & 3.65   &  1.0*         & 0.427         &    0.43 &  .36* & 2.9      \\
\hline		
		\end{tabular}
	\caption{Parameters of the nonlinear dielectric response. Values with an asterisk are assumed values.}
	\label{tab2:nonlin}
\end{table}

\subsection{Glycerol}

Fig. 1 (a) shows the fit of the linear response of glycerol at 204 K. The parameters of the fit of the linear response in terms of the pragmatical model \cite{visco} are given in Table I.

The fit of the nonlinear response at the frequency $\omega$ is shown in Fig. 1 (b) and requires a ratio $\Delta\mu/\mu=5.1$ (see Table II). Also, as already stated in the previous section, it requires the assumption of a decrease of the nonlinear strength $\tau_c/\tau^\gamma$ with $\gamma=0.2$. Otherwise the drop of the nonlinear signal toward higher frequencies is decidedly slower than the one found in experiment. The same holds for the propylene carbonate data in Figs. 3 (b) and (c), which require $\gamma=0.23$.

The value $f_3=0.72$ for glycerol in Table II was taken from a static measurement of the nonlinear effect (a reduction of $\Delta\chi$ by 0.45 percent at a field of 225 $kV/cm$) mentioned in reference \cite{r2016}.

Fig. 1 (b) shows as a dashed line the contribution from the retardation processes alone. One sees that most of the nonlinear $\omega$ effect is due to the retardation. 

\begin{figure}[b]
\hspace{-0cm} \vspace{0cm} \epsfig{file=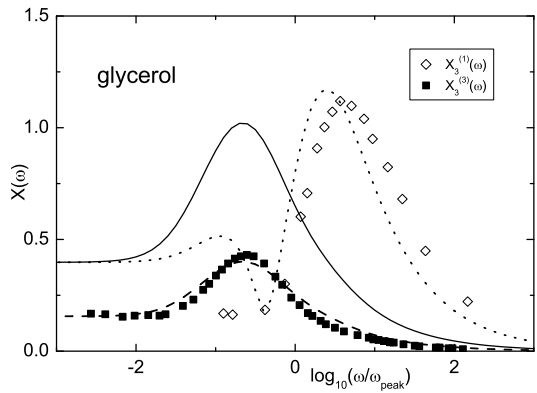,width=7 cm,angle=0} \vspace{0cm} \caption{Comparison of the model to nonlinear glycerol data \cite{brun2} at 204.7 K. The $X_1(\omega)$-data are well described with the parameters fitted to the nonlinear dielectric data at the frequency $\omega$ \cite{bauer1} in Fig. 1, but the $X_3(\omega)$-data require a reduction by a factor 2.7.}
\end{figure}

The same parameters describe the nonlinear glycerol data \cite{brun2} at the frequency $\omega$ in Fig. 2, plotted in the dimensionless form  $X_1(\omega)=NkT\chi_3^{(1)}(\omega)/\epsilon_0\Delta\chi^2$, in which they should extrapolate to $f_3f_{On}/5$ at $\omega$ equal zero.

The problem is that the data do not do what they should. This was already noted in the first evaluation \cite{brun3} in terms of extended structural rearrangements, which found the height of the hump in $\chi_3^{(3)}(\omega)$ a factor of three lower than the expectation from the $\chi_3^{(1)}(\omega)$-data.

The present analysis confirms this result and uncovers another inconsistency which may be related to the first one: the data behave as if $f_3$ was a factor of 2.7 lower than the value 0.72 extracted from the static measurement, a discrepancy with the zero frequency $3\omega$-data which was already noted there \cite{r2016}.

The fact that not only the hump in $\chi_3^{(3)}(\omega)$, but also the zero frequency limit is reduced by a factor of about three indicates a mechanism which affects $\chi_3^{(3)}(\omega)$ equally at all frequencies. We will come back to this point in the discussion.

If one forgets the absolute intensity and uses only the peak height, it is necessary to assume an asymmetry parameter $d_2=0.42$ (see Table II) (an increasing $d_2$ increases the nonlinear $\omega$-effect and decreases the nonlinear $3\omega$-effect). The value 0.42 is higher than the expectation of 1/3 for a constant asymmetry distribution.

\subsection{Propylene carbonate}

\begin{figure}[b]
\hspace{-0cm} \vspace{0cm} \epsfig{file=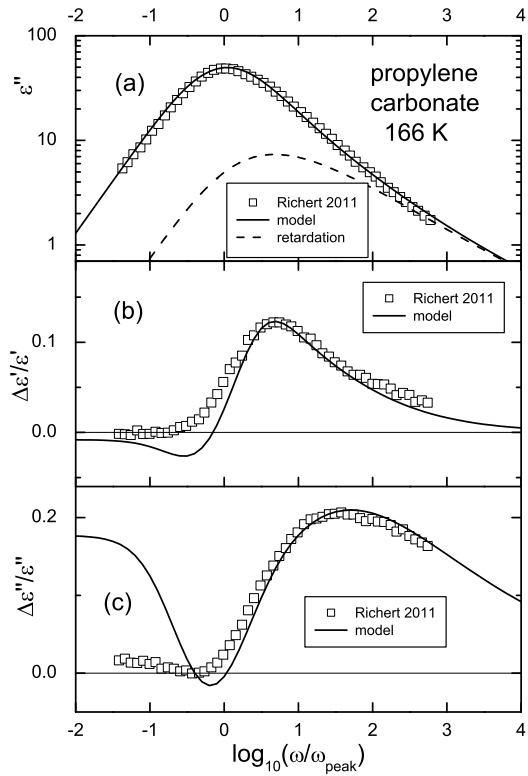,width=7 cm,angle=0} \vspace{0cm} \caption{Comparison of the model to propylene carbonate data \cite{huang,tca} (a) linear dielectric response (b) nonlinear dielectric $\Delta\epsilon'/\epsilon'$-data (c) nonlinear dielectric $\Delta\epsilon''/\epsilon''$-data at the frequency $\omega$.}
\end{figure}

Fig. 3 shows linear and nonlinear data \cite{huang,tca} of propylene carbonate at 166 K. The linear response in Fig. 3 (a) is fitted in terms of the pragmatical model with the parameters in Table I.

The $\Delta\epsilon'/\epsilon'$-data in Fig. 3 (b) were calculated from the $\Delta\tan{\delta}/\tan{\delta}$ and the $\Delta\epsilon''/\epsilon''$-data in reference \cite{huang} with the relation
\begin{equation}
	1+\frac{\Delta\epsilon'}{\epsilon'}=\frac{1+\Delta\epsilon''/\epsilon''}{1+\Delta\tan{\delta}/\tan{\delta}}.
\end{equation}

In this case, the average asymmetry $d_2=0.35$ in Table II could be fitted from the ratio of real and imaginary parts of $\chi_3^{(1)}$, which depends on $d_2$.

But the same parameters give a very poor fit of the $\chi_3^{(3)}$-data \cite{bauer3}. Transforming the data into $X_3(\omega)$, one finds again that they extrapolate to about $1/5$, so it seems again appropriate to divide the calculation by the Onsager factor. But even then, the calculated peak is a factor four higher than the data.

\subsection{Two mono alcohols}

\begin{figure}[b]
\hspace{-0cm} \vspace{0cm} \epsfig{file=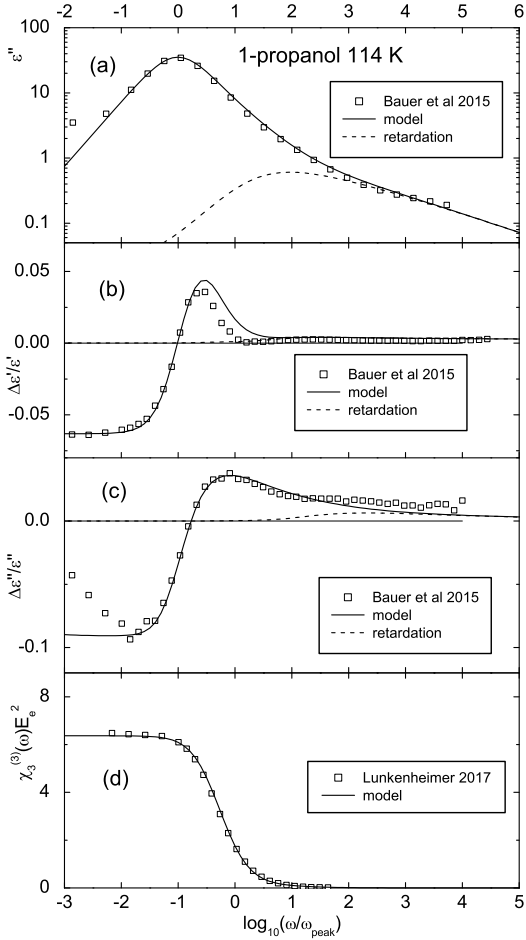,width=7 cm,angle=0} \vspace{0cm} \caption{Fit of linear and nonlinear dielectric response data of the mono alcohol 1-propanol \cite{bauer2015} in terms of the pragmatical model (a) linear dielectric response (b) $\Delta\epsilon'/\epsilon'$ (c) $\Delta\epsilon''/\epsilon''$ and (d) $\chi_3^{(3)}E_e^2$ with the same parameters.}
\end{figure}

Fig. 4 shows linear and nonlinear responses of the mono alcohol 1-propanol \cite{bauer2015,bauerdr,lunk}, .

In the fit of the linear response of Fig. 4 (a), it turned out to be necessary to include a secondary relaxation Cole-Cole peak with an amplitude of $a_{CC}=0.059$ and $\delta=0.25$. Its relaxation time was calculated from $\tau=\tau_0\exp(V/kT)$, with $\tau_0=10^{-13}$ s and $V=0.22$ eV. The other parameters are listed in Table I.

In view of the heavy consistency problems between $\omega$ and $3\omega$-effects found in glycerol and propylene carbonate, it is gratifying to find full consistency here. All three effects in Fig. 4 (b), (c), and (d) are well described with the same $f_3$-value of 1.81, which is not too far from the value 1.5 found for 1-propanol from static measurements \cite{parry}.

The nonlinear effect of the structural rearrangements around the structural lifetime $\tau_c$ is relatively weak, $\Delta\mu/\mu=2.1$. This is what one expects; the molecules keep hanging together for a much longer time and are not free to reorient completely \cite{mono}. 

The nonlinear effect $F_D$ at the Debye line is practically the same as the one for a diffusive Debye process, so there is no hump in $\chi_3^{(3)}(\omega$; it could as well be described as a diffusive Debye process of independent molecules \cite{coffey}.

\begin{figure}[b]
\hspace{-0cm} \vspace{0cm} \epsfig{file=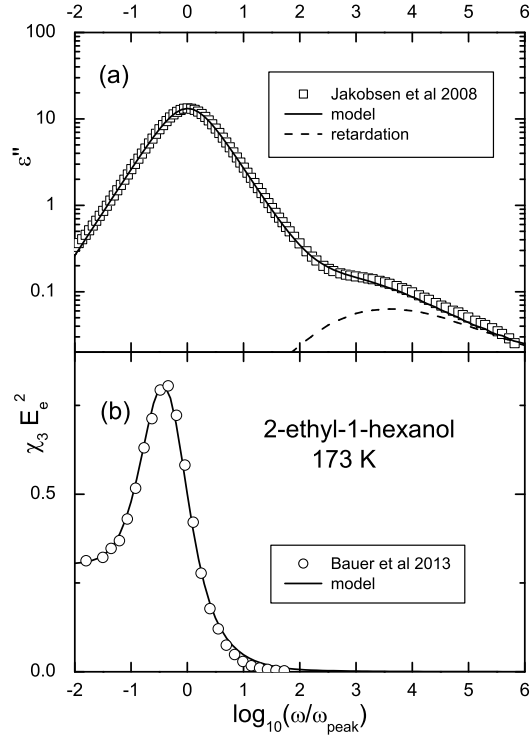,width=7 cm,angle=0} \vspace{0cm} \caption{Fit of linear \cite{bo} and nonlinear \cite{bauer3} dielectric response data of the mono alcohol 2-ethyl-1-hexanol in terms of the pragmatical model (a) linear dielectric response (b) $3\omega$-signal $\chi_3(\omega)E_e^2$.}
\end{figure}

This is different for the next and last case. Fig. 5 shows linear and nonlinear data \cite{bauer3} for a second mono alcohol, 2-ethyl-1-hexanol. The linear dielectric response in Fig. 5 (a) is fitted with the parameters in Table I, including a secondary relaxation Cole-Cole peak with an amplitude of $a_{CC}=0.041$ and $\delta=0.25$. Its relaxation time was calculated from $\tau=\tau_0\exp(V/kT)$, with $\tau_0=10^{-13}$ s and $V=0.232$ eV.

Unlike 1-propanol, 2-ethyl-1-hexanol does have a clear hump in the $3\omega$-signal \cite{bauer3}, very well fitted with $F_D=2.94$ (see Table II). The second fit result is $f_3=0.43$, a rather low value, which should be checked in a static measurement of the saturation.

\section{Discussion and conclusions}

\subsection{The missing Onsager factor}

Let us start the discussion with the $X_3(\omega)$-results at the low frequency end in glycerol (Fig. 2) and in propylene carbonate. In these two cases, one finds values which are a factor of three smaller than the expectation $f_3f_{On}/5$ of eq. (\ref{x30}) in Section II. A.

In the case of glycerol, there has been a recent determination \cite{r2016} of the decrease of the static $\Delta\chi$ by 0.45 percent in a field of 225 kV/cm, yielding $f_3=0.72$. This is a bit smaller than values in other substances \cite{parry}.

But even taking this $f_3$ into account, the zero frequency $3\omega$-signal is by about the Onsager factor of 2.7 smaller than expected (this was also noted in reference \cite{r2016}, where it is a factor of two). This is surprising; the third order effect $f_3$ can vary from substance to substance, but it should vary in the same way for $\chi_3^{(1)}$, $\chi_3^{(3)}$ and the effect in a static field.

One can pursue the effect further by comparing the prediction of eq. (\ref{x50}) to the recently measured \cite{chi5} fifth order effect in glycerol and propylene carbonate. In glycerol, one has at 204 K a fifth order change of the susceptibility of $\Delta\epsilon/E^4=4.4\ 10^{-33}$ m$^4$/V$^4$. With $\Delta\chi=63$, this translates into $X_5(0)=0.127$, again much smaller than the predicted value $2f_{On}^2/7$ ($f_{On}=2.76$).

This has again to be corrected for the factor $f_5$. $f_3$ was 0.72, so we expect a factor 0.72$^2$ for the fifth order. Taking this into account, the effect is again similar to what one finds in the third order: The signal corresponds to an Onsager factor of one. Again, something similar happens in propylene carbonate.

Note that the mechanism behind this discrepancy is not necessarily connected with the Onsager factor. From the discussion in Section II A, it is clear that the strength of the nonlinear signals depends not only on the ratio between internal and external field, but also on the interaction of the local dipole with its neighbors. In fact, one does not see how the Onsager factor could be different for $\chi_3^{(3)}(\omega)$ and $\chi_3^{(1)}(\omega)$.

The inconsistency of the $3\omega$-effect with the nonlinear $\omega$-effect is not found in the mono alcohol 1-propanol, where both effects are satisfactorily described with the same model parameters.

To get more information about this discrepancy, one should make a dedicated experiment to study the low frequency nonlinear $\omega$-signal in glycerol, to see whether $X_1(\omega)$ goes to $f_3/5$ or to $f_3f_{On}/5$ (the data in ref. \cite{brun2} seem to go to $f_3/5$, but there are only three points).

Another possible experiment is to measure the $3\omega$-signal in propylene glycol, where one knows \cite{weinstein} that the nonlinear $\omega$-signal does contain the Onsager factor.

\subsection{The decrease of the nonlinear signal at high frequencies}

In glycerol (Fig. 1 (b)) and in propylene carbonate (Fig. 3 (b) and (c)), a good fit requires a slight decrease of the nonlinear strength with decreasing retardation relaxation time $\tau_r$, describable with a power law $\tau_r^\gamma$, with $\gamma=0.2$ for glycerol and $\gamma=0.23$ in propylene carbonate.

Though the decrease is slight, it amounts to a factor 2.5 over the two decades where it is seen, clearly out of the error bars. If one attributes the strength of the nonlinear effect to the number $N_{corr}$ of molecules in a cooperatively rearranging region, this number decreases by the same factor over two decades.

But if one believes the Adam-Gibbs conjecture of a proportionality of the energy barrier to $N_{corr}$, one gets into problems. The effect is seen in the kHz region, at energy barriers of the order of 23 $kT$. Moving two decades should only change the barrier by 4.6 $kT$, i.e. change $N_{corr}$ by 20 percent, much less than a factor of 2.5.

A better explanation is the interaction between different structural rearrangements, which makes them more and more effective as one approaches the final breakdown of the shear rigidity.

\subsection{The average asymmetry}

In glycerol (see Fig. 1), one can adapt the relative height of the hump in $3\omega$ by a suitable choice of the average asymmetry $d_2$ of the two minima of the cooperatively rearranging regions. The higher the asymmetry, the lower the hump in $3\omega$ for a given $a$.

If one accepts the missing Onsager factor in the $3\omega$-signal of glycerol as a fact, one derives $d_2=0.42$, the value in Table I.

In propylene carbonate, one can fit $d_2$ to the ratio between real and imaginary parts of $\chi_3^{(1)}(\omega)$, and gets 0.35, again a value larger than the expectation value of 1/3 for a constant distribution of asymmetries.

As explained in Section II. B , $d_2=1/3$ is the crucial point where the contribution of a cooperatively rearranging region to the low frequency susceptibility changes sign. At lower $d_2$, the saturation dominates and the contribution is negative, both in $\omega$ and $3\omega$. At $d_2>1/3$, the population increase of the upper minimum in the strong alternating electric field dominates and the contribution is positive.

Since the whole low frequency susceptibility must be negative to obey the Langevin saturation (see II.A) a positive contribution from the cooperatively rearranging regions has to be overcompensated by the viscous contribution according to eq. (\ref{deta}). This means that $d_\eta$ has to be smaller than $1/3$.

\begin{figure}[b]
\hspace{-0cm} \vspace{0cm} \epsfig{file=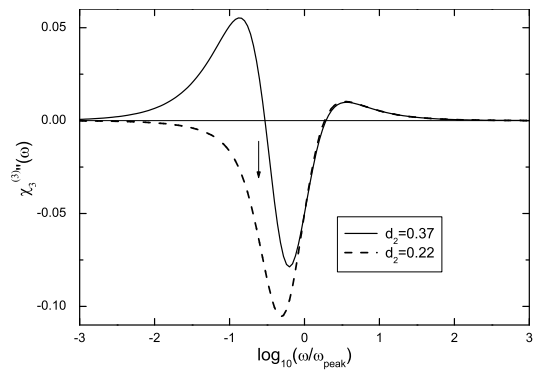,width=7 cm,angle=0} \vspace{0cm} \caption{
Retardation (continuous line) and viscous (dashed line) components of the imaginary part of the hump in $3\omega$, denoted by the arrow, in glycerol. Note the negative interference at and below the hump.}
\end{figure}

Fig. 6 illustrates the mechanism for the decrease of the $3\omega$-hump for the glycerol case. It is mainly due to the imaginary part of $\chi_3^{(3)}(\omega)$, which is the sum of a retardation part with $d_2=0.37$ and a viscous part with $d_\eta=0.23$. One can see that there is a negative interference at and below the arrow which denotes the position of the hump. Something similar happens in the real component.

Since the $\chi_3^{(1)}$-component is dominated by the retardation (see Fig. 1 (b)), it is much less affected by this negative interference.

\subsection{The value of $N_{corr}$}

Let us address the question whether one can determine the number $N_{corr}$ of molecules in a cooperatively rearranging region from the ratio $\Delta\mu/\mu$.

One can calculate a lower bound for $N_{corr}$ by assuming that each molecule in a cooperatively rearranging region changes its orientation in a completely statistical way. This leads to a contribution of $2\mu^2/3$ per molecule to the total mean squared polarization difference, so
\begin{equation}
	N_{corr,lb}=\frac{3\Delta\mu^2}{2\mu^2},
\end{equation}
which in the case of glycerol leads to an $N_{corr}$-value of 33 molecules.

But in glycerol, one is in the happy situation of knowing the jump angle distribution in the flow process from NMR data \cite{bohmer}. One finds a bimodal distribution, with many small-angle jumps and a few large-angle ones, with a rotation angle of 30 degrees. It seems reasonable to associate the small-angle jumps with molecules outside the cooperatively rearranging region and the 30-degree jumps with the molecules inside.

A jump of 30 degrees leads to a $\Delta\mu=0.518\mu$, so now one has only a contribution of $0.268\mu^2$ per molecule to the total mean squared polarization difference, less than half of the value for completely statistical reorientation.

The NMR investigation \cite{bohmer} finds a similar result for toluene, a bimodal distribution with a large jump angle of 25 degrees, implying that a smaller jump angle than the statistical one does not only occur in hydrogen bonded substances, but also in van-der-Waals bonded ones.

From the NMR results, it seems more realistic to assume a rule of thumb
\begin{equation}\label{ncorr}
	N_{corr}\approx 4\frac{\Delta\mu^2}{\mu^2},
\end{equation}
which in glycerol leads to about ninety molecules in a cooperatively rearranging region, a reasonable number. In propylene carbonate, the recipe gives one hundred and sixty molecules.

In this context, it should be noted that the new version of the pragmatical model \cite{visco} is in better agreement with the NMR results \cite{bohmer} than the first one \cite{visc}. In the old one, all of the response was due to structural rearrangements; in the new one, the larger part $f_c$ of the response is viscous diffusive angular motion due to a collective effect of the structural rearrangements. This is in fact an important part of the NMR result \cite{bohmer}, namely that a larger part of the total response is small angle motion.

\subsection{Mono alcohols}

In mono alcohols, the chain of connected molecules is like a long worm transversing the cooperatively rearranging region \cite{mono}, a worm which survives hundreds of rearrangements. The surviving chain connection must considerably reduce the mean squared polarization difference of the rearrangement. 

This is consistent with the value $\Delta\mu/\mu=2.1$ found in 1-propanol (see Table II), decidedly smaller than the values 5.1 and 8.1 in glycerol and propylene carbonate. The cooperatively rearranging region might have the same size in glycerol and in 1-propanol, but the resulting dipole moment change would still be smaller, because neighboring molecules keep their connection.

Without such a connection, the ratio $\Delta\mu/\mu$ allows a fair estimate of $N_{corr}$ with the help of eq. (\ref{ncorr}). But this is not the case for the ratio $F_D$ measured at the Debye line of a mono alcohol. $F_D$ is certainly a measure of the cooperativity of the Debye process, but remember that it has been obtained with a completely inadequate theoretical picture (in fact, it is astonishing that it provides such a good fit for 2-ethyl-1-hexanol in Fig. 5 (b)). The Debye line is certainly not a jump process, but a diffusional process. To obtain its cooperativity, it must be fitted in an appropriate diffusional picture.

\subsection{Comparison to the box model}

As stated in the introduction, the box model \cite{weinstein,tca} describes the $\omega$-effect in terms of an acceleration of the rearrangements, due to the heating by the strong electric field.

\begin{figure}[t]
\hspace{-0cm} \vspace{0cm} \epsfig{file=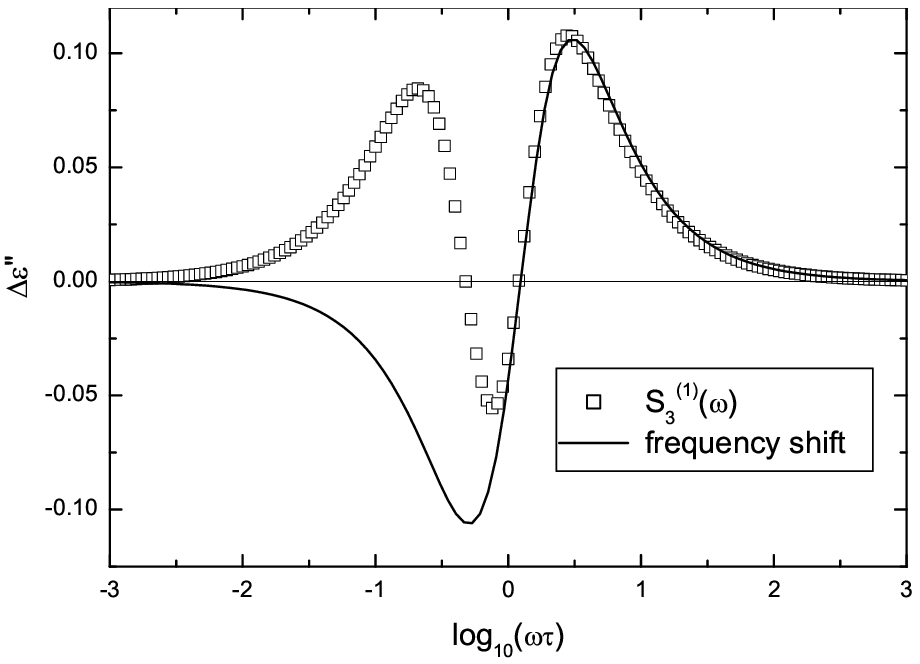,width=7 cm,angle=0} \vspace{0cm} \caption{Comparison of the nonlinear imaginary $\omega$-response of an asymmetric double-well potential with the asymmetry $d_2=0.37$ (the average value in glycerol and propylene carbonate) to the pure acceleration response postulated in the box model \cite{weinstein,tca}.}
\end{figure}

Fig. 7 compares the two theoretical concepts in their effect on the quantity $\Delta\epsilon''$ for a single cooperative rearrangement with $d_2=0.37$, close to the average values found in glycerol and propylene carbonate. It is seen that the effects are similar at frequencies higher than the inverse relaxation time, but very different at lower frequencies.

This is easy to understand. The effect of a strong alternating electric field on the thermally activated motion in the asymmetric double-well potential \cite{brun,gregor} is dominated at high frequency by the increase of the rates (see eq. (\ref{rate})). There, both concepts are in accordance.

But at low frequency, saturation and the average population increase of the higher well begin to dominate the response in the double-well picture. The corresponding response is very different from a simple acceleration of the mode, not only for $d_2=0.37$, but for any asymmetry.  

Quantitatively, a value $E\Delta\mu/kT=1$ leads to an acceleration factor 1.066 at high frequency. The fitted values for glycerol and propylene carbonate in Table II essentially provide the same acceleration as the box model ones calculated from the excess heat capacities and from the temperature dependence of the $\alpha$-relaxation times \cite{weinstein,tca}.

In the cooperative rearrangement picture, one can understand \cite{brun4} the surprising success of the box model qualitatively (but not quantitatively) by invoking the proportionality of $\partial\ln\tau_\alpha/\partial\ln T$ to $N_{corr}$. As was seen in the discussion of the $N_{corr}$-value above, the relation between $N_{corr}$ and $\Delta\mu$ involves an average jump angle, which is not {\it a priori} fixed. Therefore the quantitative success of the box model \cite{weinstein,tca} appears fortuitous from the point of view of the cooperativity model. 

In fact, the box model prediction for the $3\omega$-hump fails to give quantitative agreement \cite{brun4}. In order to get a successful fit \cite{r2016} of new glycerol $3\omega$-data, it is necessary to invoke not only the box model effect (called "energy absorption mechanism" in the paper), but a field induced entropy reduction and a saturation effect as well.

\subsection{Conclusions}

To conclude, glycerol and propylene carbonate show a surprising absence of the Onsager factor in their $3\omega$ and $5\omega$-response. The theoretical reason is completely unclear, the more so as the mono alcohol 1-propanol has the Onsager factor in its $3\omega$-response.  

The pragmatical model, which allows to separate the viscous response from the retardation response of thermally activated back-and-forth jumps, has been extended to describe the nonlinear dielectric response in molecular glass formers. The retardation response is calculated from Diezemann's equations for the nonlinear response of asymmetric double-well potentials.

The nonlinear response of the viscous component is described in terms of a single asymmetric double-well potential. The relaxation time of this single asymmetric double-well potential is supposed to lie at the terminal dielectric relaxation time, which in simple molecular glass formers is identical with the lifetime of the retardation potentials, but which is a factor of the order of hundred longer in the mono alcohols.

In the pragmatical model, the viscous no-return jumps occur in the retardation potentials at the end of their lifetime. Therefore the viscous response should exhibit the same average polarization difference as the retardation jumps, while the asymmetry of the viscous double-well potential must be chosen to arrive at the correct single-molecule Onsager low-frequency limit.

The model gives a successful description of the measured nonlinear spectra at the frequency $\omega$ of the strong alternating electric field. The description requires a slightly larger average asymmetry of the double-well potentials than the one for a constant distribution of asymmetries. 

While the mono alcohol 1-propanol shows practically no cooperativity effects at its dominating Debye component, the mono alcohol  2-ethyl-1-hexanol exhibits a clear hump in the $3\omega$ effect, understandable in terms of a transition from a motion of larger units at frequencies above the chain-breaking Debye line to the motion of single molecules at lower frequencies.

Thanks for very valuable discussions are due to Ranko Richert, Alois Loidl, Peter Lunkenheimer (in this case also for sending data of references \cite{bauer3,bauer2015}) and Francois Ladieu.

\end{document}